\documentclass{webofc}
\usepackage[varg]{txfonts}   % Web of Conferences font
\usepackage{hyperref}
\usepackage{url}
\usepackage{amsmath,bm}

\def\thalf{{\textstyle{\frac{1}{2}}}}

\hypersetup{colorlinks=true,citecolor=blue,urlcolor=blue,linkcolor=blue}
\begin{document}
\title{Spinodal decomposition in Bjorken flow}
%
% subtitle is optionnal
%
%%%\subtitle{Do you have a subtitle?\\ If so, write it here}

\author{\firstname{Joseph} \lastname{Kapusta}\inst{1}\fnsep\thanks{\email{kapusta@umn.edu}} \and
        \firstname{Mayank} \lastname{Singh}\inst{1,2}\fnsep\thanks{\email{mayank.singh@vanderbilt.edu}} \and
        \firstname{Thomas} \lastname{Welle}\inst{1,3}\fnsep\thanks{\email{twelle2357@gmail.com}}
        % etc.
}

\institute{School of Physics and Astronomy, University of Minnesota, Minneapolis, MN 55455, USA
\and
           Department of Physics and Astronomy, Vanderbilt University, Nashville, TN 37240, USA
\and
           Applied Research Associates, 8537 Six Forks Road, Raleigh, NC 27615, USA
          }

\abstract{The QCD first-order phase transition at large baryon densities is expected to proceed by spinodal decomposition. This spinodal phase is likely to leave its signatures on the experimental observables measured in heavy-ion collision experiments. Identifying these signatures requires phenomenological models integrating surface effects resulting from the phase transition into the hydrodynamical description of the expanding quark gluon plasma. We write the equations of relativistic hydrodynamics with spinodal decomposition and solve it in on a background of Bjorken flow relevant for heavy-ion collisions.
}
\maketitle
\section{Introduction}

Mapping the QCD phase diagram has been a priority for the heavy-ion collision community. Lattice calculations with non-zero quark masses have shown that the QCD matter changes phase from a hadron gas to the quark gluon plasma (QGP) by a smooth crossover when the baryon chemical potential is near zero \cite{Aoki:2006we}. At high baryon densities, the phase change is expected to proceed via a first-order phase transition \cite{Fukushima:2010bq}. The first-order phase transition curve ends in a critical point before reaching the zero chemical potential axis.

The Beam Energy Scan (BES) program at the Relativistic Heavy-Ion Collider (RHIC) and the upcoming experiments at the Facility for Antiproton and Ion Research (FAIR) aim to explore different regions of the QCD phase diagram with the goal of quantitatively mapping its features. Heavy nuclei are collided at a range of collision energies which cover a wide region of the phase diagram. This program relies on identifying the observable signatures of the critical point and the phase transition curve and doing extensive model-to-data comparisons \cite{An:2021wof}. We expect the phase transition curve to cover a bigger region of the phase diagram than the critical point and hence it is crucial that we understand its dynamics.

The phase transition in these collisions is expected to proceed via spinodal decomposition. Here we extend the equation of state to metastable and unstable regions and report on the equations describing spinodal decomposition in a relativistic QCD medium. We write and solve these equations in a fluid with Bjorken flow.

\section{Metastable and unstable regions}

\begin{figure}
    \centering
    \includegraphics[width=0.45\linewidth]{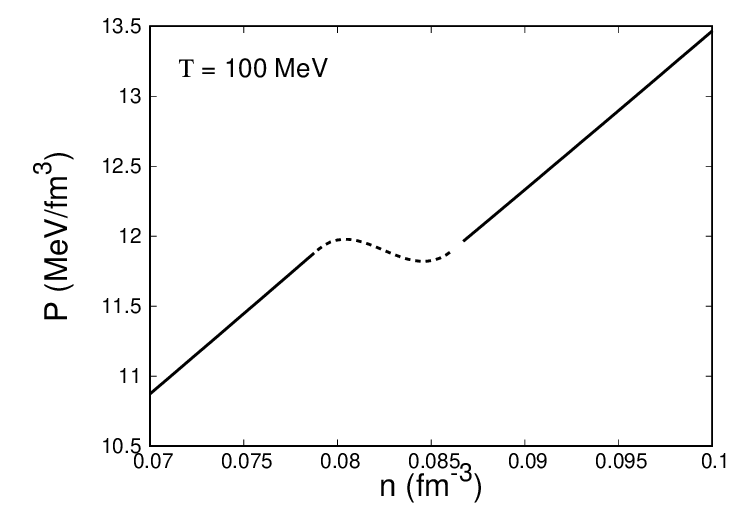}
    \includegraphics[width=0.45\linewidth]{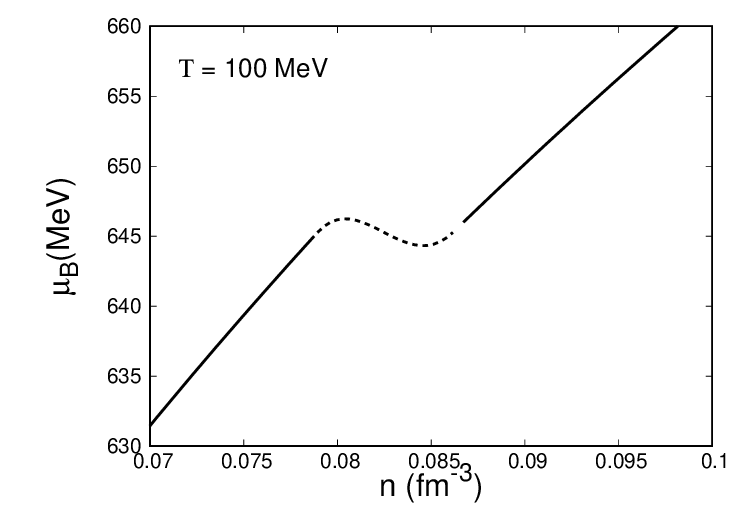}
    \caption{Pressure (left) and baryon chemical potential (right) are interpolated in metastable and unstable phases at fixed temperature.}
    \label{fig1}
\end{figure}

An important step towards describing the spinodal decomposition is extending the equation of state to the coexistence phase. This phase is often not described by the first principle approaches as it is composed of metastable and unstable regions. Given an equation of state for the stable region, we can interpolate the pressure in this phase as \cite{Kapusta:2024nii}
\begin{equation}
P_{{\rm int}}(n) = P_X(T) + \sum_{i = 1}^4 c_i (n - n_G)^i.
\end{equation}
The subscript $X$ denotes the pressure along the line where the phases coexist. This line is often defined by Maxwell construction. The baryon density is denoted by $n$ while $n_G$ denotes the baryon density in the hadronic gas phase along the coexistence line. Similarly, the chemical potential can be parameterized as
\begin{equation}
\mu_{{\rm int}}(n) = \mu_X(T) + d_0 \ln(n/n_G) + \sum_{i = 1}^3 d_i (n-n_G)^i,
\end{equation}
The coefficients $c_i$ and $d_i$ can be obtained by applying the thermodynamic relations at $n = n_G$ and $n = n_L$ where the equation of state is stable. Further constraints are obtained by applying the continuity conditions on pressure and chemical potential and their derivatives on the border between the coexistence and the QGP and hadronic gas phases. The thermodynamic relations also hold at the boundaries and provide additional constraints. The interpolated pressure and baryon chemical potential for a temperature of 100 MeV are shown in figure \ref{fig1}. The stable QCD equation of state is taken from reference \cite{Kapusta:2021oco}.

\section{Hydrodynamics with phase separation}
The Helmholtz free energy can be written as
\begin{equation}
F\{n({\bf x},t)\} = \int d^3x \left[ \thalf K (\bm{\nabla} n)^2 + f(T,n) \right],
\end{equation}
where first function on the right accounts for the free energy due to a gradient in baryon density across the phase boundary. The second term $f(T,n)$ is the bulk free energy. Cahn and Hilliard \cite{CahnHilliard1,CahnHilliard2} showed in their seminal papers that this leads to the stress tensor
\begin{equation}
T_{ij} = \tilde{P} \delta_{ij} + K (\partial_i n) (\partial_j n).
\end{equation}
For a relativistic fluid, the covariant energy-momentum tensor can be given by \cite{Kapusta:2024nii}
\begin{equation}
T^{\mu\nu} = \tilde{P} (u^{\mu} u^{\nu} - g^{\mu\nu}) + \tilde{\epsilon} u^{\mu} u^{\nu} + K (D^{\mu} n) (D^{\nu} n),
\end{equation}
where the local thermodynamic variables are pressure $\tilde{P} = P + K n D^2 n + \thalf K (D^{\mu} n) (D_{\mu} n)$ and energy density $\tilde{\epsilon} = \epsilon - \thalf K (D^{\mu} n) (D_{\mu} n)$. Here $D^\mu = \partial^\mu - u^\mu u^\alpha \partial_\alpha$ is the gradient orthogonal to the fluid velocity $u^\mu$. We are using the mostly negative metric. In the Landau-Lifshitz frame baryon current has the form
\begin{equation}
J^{\mu} = n u^{\mu} + \sigma_B T D^{\mu} \left( \frac{\tilde{\mu}}{T} \right),
\end{equation}
with baryon chemical potential $\tilde{\mu} = \mu + K D^2 n$ and baryon conductivity $\sigma_B$.

We solve equations for boost invariant hydrodynamics with the simplification that system is homogeneous in transverse direction. We start the evolution at proper time $\tau_x = 10$ fm when energy density is kept boost invariant. Baryon density is initialized as $n(\tau_x) = n_0(\tau_x) + n_{fl} \sin(k \xi)$ where $n_0$ is the average value of baryon density. Here $n_{fl}$ is the amplitude of sinusoidal fluctuations in baryon density in the rapidity direction $\xi$. The values of $n_0$ and $n_{fl}$ are chosen such that the lowest point of baryon density distribution just touches the coexistence phase region. Energy density gradient terms are neglected for simplicity. We verified that the violation of boost invariance in energy density is less than 3\%.

\begin{figure}
    \centering
    \includegraphics[width=0.45\linewidth]{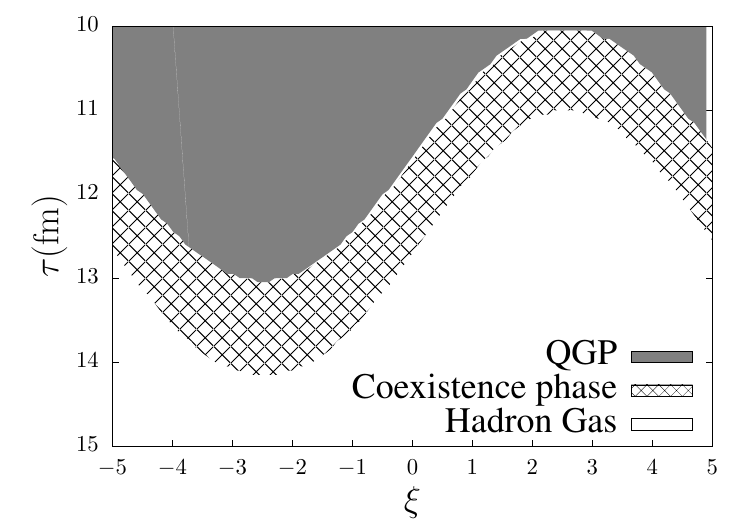}
    \includegraphics[width=0.45\linewidth]{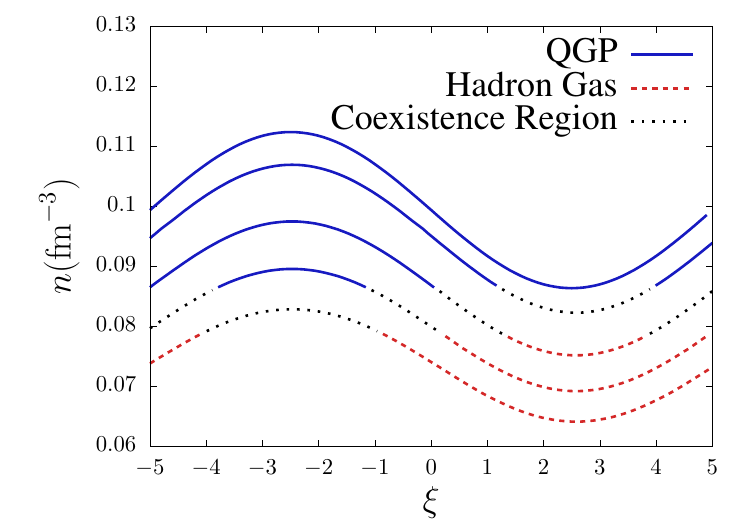}
    \caption{The phases of QCD matter as a function of proper time and rapidity (left). The spatial distribution of baryon density (right). The five curves from top to bottom correspond to $\tau =$ 10 fm, 10.5 fm, 11.5 fm, 12.5 and 13.5 fm respectively.}
    \label{fig2}
\end{figure}

The evolution of different phases and that of baryon density can be seen in figure \ref{fig2}. The baryon density reduces as the system expands and cools with time. It is interesting to note that unlike the steady-state systems undergoing spinodal decomposition, which have been extensively studied, this is a rapidly expanding system. So the system always dilutes and ends up entirely in the hadronic gas phase. If the phase change is via a first-order transition, then it encounters the coexistence phase.

The baryon density chosen here is smoothly varying. This is a simplification and ensures that the fourth-order derivatives encountered in the Cahn-Hilliard model are simplified. We chose a relatively small value of $K = 5\times10^{-5}$ MeV$^{-4}$ which ensured that this simplification holds for the duration of this evolution. In the future, this simplification can be relaxed for a more realistic baryon density distribution.

\begin{figure}
    \centering
    \includegraphics[width=0.45\linewidth]{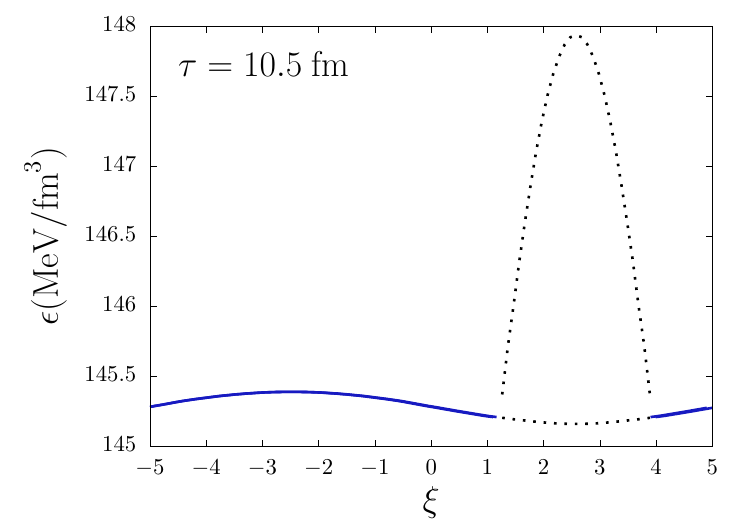}
    \includegraphics[width=0.45\linewidth]{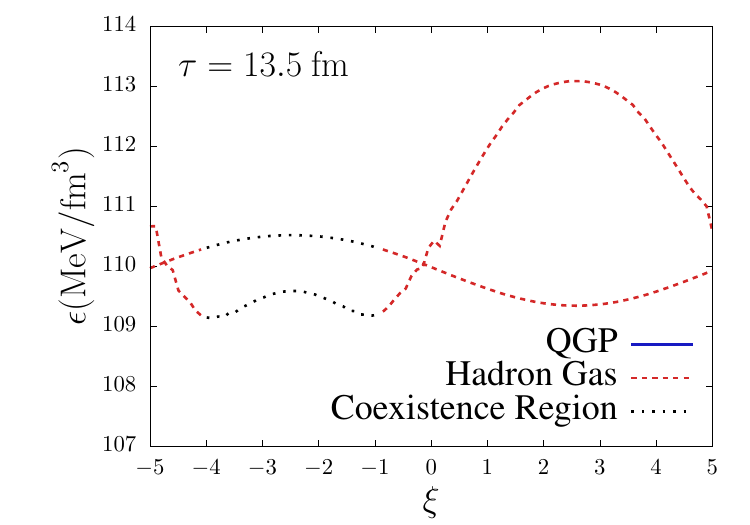}
    \caption{Energy density evolution for two different values of $\tau$. The curves with higher maximum value of $\epsilon$ correspond to $K = 5\times10^{-5}$ MeV$^{-4}$. The other curves corresponds to $K = 0$.}
    \label{fig3}
\end{figure}

The energy density evolution is shown in figure \ref{fig3}. We show the results with $K = 5\times10^{-5}$ MeV$^{-4}$ and $K = 0$. The later choice means that the phase surface has no energy contribution and is shown for reference. The energy increases immediately as the system enters the coexistence region. The energy ends up reducing in the negative rapidity when it goes in the coexistence phase. Whether the energy density increases or decreases depends on the sign of baryon density gradients. These surface energy contributions have the net effect of enhancing the fluctuations already present in the system.

\section{Conclusions}

The QCD phase diagram is postulated to have a first-order phase transition curve at large baryon densities. Experimental verification of such a phase transition relies on successful model-to-data comparisons with the models incorporating the phase transition dynamics. In heavy-ion collisions, such a phase transition is expected to proceed by spinodal decomposition. We wrote the equations of relativistic hydrodynamics with spinodal decomposition. The covariant energy-momentum tensor incorporating the surface effects is explicitly shown. We also provide an interpolation prescription to extend the QCD equation of state to metastable and unstable phases. We solved the equations for Bjorken flow.

The formalism can be included in the existing models of heavy-ion collisions to simulate the QCD phase transition. This will help us identify the observables containing the signatures of the phase transition and help map the QCD phase diagram.

\section*{Acknowledgements}
This work was supported by the U.S. Department of Energy Grant Nos. DE-FG02-87ER40328 (JK, MS and TW) and DE-SC-0024347 (MS).
% BibTeX or Biber users please use (the style is already called in the class, ensure that the "woc.bst" style is in your local directory)
% \bibliography{your_bib_file} % Replace "your_bib_file" with the actual name of your .bib file
%
% Non-BibTeX users please use
%

\end{document}